\documentclass[%
 reprint,
 amsmath,amssymb,
 aps,prc
]{revtex4-2}
\usepackage[colorlinks, citecolor=red]{hyperref}
\usepackage{graphicx}
\usepackage{dcolumn}
\usepackage{bm}
\usepackage{nameref}
\usepackage{natbib}
\usepackage[T1]{fontenc}
\usepackage{booktabs, array, mathptmx, float, tabularx, booktabs, lipsum, amsmath,multirow}
\usepackage{siunitx, xcolor}
\usepackage[version=4]{mhchem}
\usepackage{booktabs}
\usepackage{tabularx}
\usepackage{xtab}
\begin{document}
\preprint{APS/123-QED}
\title{Nonlocality Effect in the Alpha decay half-lives of superheavy nuclei with XGBRegressor}

\author{ Jinyu Hu$^{1}$ and  Chen Wu$^{1}$ } \affiliation{
\small 1. Xingzhi College, Zhejiang Normal University, Jinhua, 321004, Zhejiang, China}
\begin{abstract}
Building on the work of E. L. Medeiros \cite{medeiros2022nonlocality} and our previous study \cite{hu2025nonlocality}, we generalize the alpha-nucleus nonlocality effect to odd-A and odd-odd nuclei within the two-potential approach (TPA) framework. The coordinate-dependent parameters introduced by this nonlocality are optimized using an advanced gradient boosting regression model. This improved TPA is applied to calculate the $\alpha$-decay half-lives of 599 nuclei with $Z = 52-118$, yielding a root-mean-square (RMS) deviation that is 74.8$\%$ lower than that of the original TPA. Subsequently, we employ the improved TPA, along with the DZR \cite{deng2020improved} and MUDL \cite{soylu2021extended} models, to predict alpha-decay half-lives for 142 superheavy nuclei with $Z = 117-120$. The predictions from all three models are in close agreement, with the results from our improved TPA and the DZR model being nearly identical.
\end{abstract}

\maketitle

\section{\label{sec:level1}INTRODUCTION}
Alpha decay was first explained by Ernest Rutherford as the process in which a nucleus emits a $^{4}$He nucleus. This phenomenon, subsequently termed alpha decay, has remained a central subject of study in nuclear physics. Advances in experimental techniques \cite{ma2020short,oganessian2011eleven,oganessian2004experiments,zhang2021new,oganessian2000observation,oganessian2011synthesis} have led to the discovery of increasingly many nuclei, including synthesized superheavy nuclei. A prominent synthesis route for superheavy nuclei is the cold fusion of $^{208}$Pb with neutron-rich beams (A > 50), utilized in laboratories including GSI, RIKEN, and JINR. A significant milestone was reached through $^{48}$Ca-induced hot fusion reactions using an actinide californium target, leading to the synthesis of a nucleus with $Z = 118$  \cite{oganessian2011synthesis}. This success has extended the known boundaries of the periodic table. As the predominant decay mode for heavy and superheavy nuclei, the study of  $\alpha$-decay has grown in importance. It provides critical guidance for the synthesis and identification of superheavy nuclei. Moreover, $\alpha$-decay data yield valuable information on nuclear structure, such as low-lying states, round-state properties, energy level structures, and shell closure effects. \cite{matsuse1975study,wauters1994fine,kucuk2020role,wang2017competition,xiao2020alpha}
\par{}
The empirical study of alpha decay began with the Geiger-Nuttall law proposed in 1911, which relates the logarithm of the half-life linearly to the decay energy. This seminal contribution was followed by various refined empirical formulas, including: the VSS \cite{viola1966nuclear} and Royer formulas\cite{royer2000alpha}, which introduce a proton-number dependence; the AKRA formula \cite{akrawy2018new}, which incorporates an asymmetry term; the MYQZR formula \cite{deng2020improved}, which considers both asymmetry and angular momentum; the Universal Decay Law (UDL) \cite{qi2009universal} for general decay processes; and its enhanced form, the MUDL \cite{soylu2021extended}, which further includes angular momentum. The new empirical, which considers  nuclear deformation, was proposed by V.Y.Denisov \cite{denisov2024empirical}.
\par{}
The quantum-mechanical explanation of alpha decay was established in 1928 through the independent work of Gurney and Condon \cite{gurney1928wave}, and Gamow \cite{gamow1928quantentheorie}. Their theory describes the process as the quantum tunneling of an alpha particle through the Coulomb potential barrier. This framework provided a fundamental explanation for the empirical Geiger-Nuttall law, namely the linear relation between the logarithm of the half-life and the inverse square root of the decay energy. With the advancement of nuclear physics, a variety of refined theoretical formalisms have since been developed  \cite{buck1992alpha,duarte1998cold,goncalves1993effective,basu2003role,chowdhury2006alpha}. Such as the two-potential approach \cite{gurvitz1987decay}, density-dependent M3Y effective interaction \cite{samanta2007predictions}, cluster-formation model \cite{ahmed2017alpha}, generalized liquid drop model \cite{zhang2006alpha}, and fission-like model \cite{buck1993half}.
\par{}
In nuclear physics, machine learning (ML) techniques have been widely applied to predict various nuclear properties, including nuclear masses, charge radii, neutron drip lines, and the half-lives of $\beta$-decay and $\alpha$-decay. To further enhance predictive performance, Bayesian neural networks have also been employed to model nuclear masses, $\alpha$-decay half-lives, and charge radii. In addition, artificial neural network architectures are being actively explored in contemporary nuclear physics research. Recently, Amir Jalili \cite{jalili2024decay} et al. applied support vector machines to the study of $\alpha$-decay half-lives and demonstrated that this approach outperforms many empirical formulas. Yang et al \cite{yang2026alpha}. employed support vector regression, using the mass number, proton number, $\alpha$-decay energy, and angular momentum as input features to calculate $\alpha$-decay half-lives. They further utilized this model to predict the $\alpha$-preformation probability. Overall, support vector regression models provide a more accurate description of $\alpha$-decay half-lives than the original density-dependent folding model (DFM). Furthermore, we have applied decision tree regression, random forest regression, and gradient boosting regression to calculate the $\alpha$-decay half-lives of 198 even-even nuclei \cite{hu2025nonlocality}.
\par{}
Aiming to improve half-life predictions, the nonlocality effect of the alpha nucleus was incorporated into the TPA framework, which was applied to the $\alpha$ decay half-lives of 599 nuclei with $Z = 52 - 118$. The parameters arising from this modification were predicted via a XGBRegressor model. This machine-learning-enhanced TPA achieves a 74.85$\%$ increase in accuracy over the original model, despite a concomitant rise in the standard deviation relative to experimental values. Furthermore, the refined model was employed to predict alpha-decay half-lives for 142 superheavy nuclei with $Z=117-120$. Comparative analysis with the established DZR model proposed by Deng et al \cite{deng2020improved} and the MUDL model proposed by Soylu et al \cite{soylu2021extended}, revealed that our predictions are in close agreement with those of the DZR model.
\par{}
The remainder of this article is organized as follows. Section \ref{sec:level2} briefly introduces the theoretical framework of the two-potential approach (TPA), nonlocality effect in alpha-particle and Machine learning. The numerical results and discussion are shown in Section \ref{sec:level3}. Finally, a summary and result are given in Section \ref{sec:level4}.
\section{\label{sec:level2}THEORETICAL FRAMEWORK }
 \subsection{TPA framework}\label{A}
In this article, the half-life $T_{1/2}$ of alpha decay can be expressed as:
\begin{equation}\label{eq1}
T_{1/2}= \frac{\mathrm{\hbar ln 2}}{\Gamma},
\end{equation}
where $\hbar$ is the reduced Planck constant.
Within the framework of the two-potential approach (TPA), the $\alpha$ decay width $\Gamma$ is directly proportional to the normalization constant $F$, the $\alpha$-preformation factor $P_{\alpha}$, and the penetration probability $P$, which can be written as:
\begin{equation}\label{eq2}
\Gamma= \frac{\hbar^{2}P_{\alpha} F P}{4\mu},
\end{equation}
where $\mu$ is the reduced mass.Meanwhile, the normalization constant $F$ is determined by the integral over the internal region and can therefore be written as:
\begin{equation}\label{eq3}
F= \frac{1}{\int_{r_{1}}^{r_{2}}\frac{1}{2k(r)}dr },
\end{equation}
In Eq. (\ref{eq3}), $k(r) = \sqrt{(\frac{2\mu}{\hbar^{2}}\left | Q_{\alpha} - V(r) \right | )}$ is the wave number of the $\alpha$ particle. Here, represents the reduced mass of the $\alpha$-daughter nucleus system in the center-of-mass frame. The quantities $Q$ and $V(r)$ denote the $\alpha$-decay energy and the $\alpha$-core potential, respectively. Finally, within the WKB approximation, the penetration probability $P$ can be expressed as:
\begin{equation}\label{eq4}
P=\mathrm{exp} \left [  -2 \int_{r_{2}}^{r_{3}}k(r)dr  \right ],
\end{equation}
In the internal region ($r_{1}$ < $r$ < $r_{2}$), the system is in the $\alpha$-preformation state, which is governed by the nuclear interaction within the $\alpha$-daughter nucleus system. In the external region ($r_{2}$ < $r$ < $r_{3}$), the Coulomb interaction acts as a potential barrier and determines the penetration probability of the $\alpha$-particle. Accordingly, the points $r_{1}$, $r_{2}$, and $r_{3}$  denote the classical turning points at which the potential energy equals the $\alpha$-decay energy.
\par{}
In this framework, the total $\alpha$-core potential energy can be expressed as the sum of three contributions: the nuclear potential energy $V_{N}(r)$, the Coulomb potential energy $V_{C}(r)$, and the centrifugal potential energy $V_{l}(r)$. It can be written as:
\begin{equation}\label{eq5}
V(r)=V_{N}(r)+V_{C}(r)+V_{l}(r).
\end{equation}
\par{}
Based on the analysis of $\alpha$-decay data, we employ a type of cosh parametrized form to describ the nuclear potential, which takes the following form:
\begin{equation}\label{eq6}
V_{N}(r)=-V_{0}\frac{1+\mathrm{cosh(R/a)}}{\mathrm{cosh(r/a)}+\mathrm{cosh(R/a)}},
\end{equation}
here, $V_{0}$ and $a$ denote parameters of the depth and diffuseness of the nuclear potential. 
The two parameters are defined as follows. Parameter $a$ is fixed at 0.5958 fm. Parameter $V_{0}$  adopts the form of a linear function in the isospin \cite{sun2016systematic} ($V_{0} = 192.42 +31.059 \frac{N-Z}{A}$), where $A$, $Z$, and $N$ represent the mass number, proton number, and neutron number of the daughter nucleus, respectively. Under the assumption of a uniformly charged spherical nucleus, the Coulomb energy $V_{C}(r)$ can be expressed as:
\begin{equation}\label{eq7}
V_{C}(r)  =\left\{\begin{matrix}
 \frac{Z_{d}Z_{\alpha}e^{2}}{2R}\left [  3 - \frac{r^{2}}{R^{2}}\right ] & r\le R \\
  \frac{Z_{d}Z_{\alpha}e^{2}}{r},& r > R,
\end{matrix}\right.
\end{equation}
where $Z_{d}$ is the charge number of the daughter nucleus, $Z_{\alpha}$ is the charge number of the $\alpha$-particle.
$R$ denotes the sharp radius of interaction, it is written as:
\begin{equation}\label{eq8}
R=1.28 A^{1/3} - 0.76+ 0.8A^{-1/3}.
\end{equation}
\par{}
In the case of unfavored $\alpha$-decay, the centrifugal potential must be taken into account. It is determined by the angular momentum of the emitted $\alpha$-particle and can be written as:
\begin{equation}\label{eq9}
V_{l}(r)=\frac{\hbar^{2}(l+\frac{1}{2})^{2}}{2\mu r^{2}}.
\end{equation}
\par{}
For one-dimensional problems, the centrifugal potential is modified as follows: $l(l+1)$ $\to $ $(l+\frac{1}{2})^{2}$.
Based on the conservation of spin-parity, the minimum value of the angular momentum $l_{min}$ that the emitted $\alpha$-particle must carry can be determined as:
 by \cite{sun2017systematic}
\begin{equation}\label{eq10}
l_{min}= \left\{\begin{matrix}
 \Delta_{j}, &for  &even-even  & \Delta_{j} &and  &\pi_{p}=\pi_{d}, \\
  \Delta_{j+1},& for &even-odd  & \Delta_{j} &and  & \pi_{p}\ne \pi_{d},\\
  \Delta_{j},&  for&  odd-even& \Delta_{j} & and & \pi_{p}=\pi_{d},\\
  \Delta_{j+1},&  for&  odd-even& \Delta_{j} & and &\pi_{p}\ne \pi_{d},
\end{matrix}\right.
\end{equation}
where $\Delta_{j} = \left | j_{p} - j_{d} \right |$, $j_{p}$, $\pi_{p}$, $j_{d}$, $\pi_{d}$ denote the spin and parity values of parent and daughter nuclei, respectively \cite{sun2017systematic}.
\par{}
This study adopts the cluster formation model (CFM). Originally proposed by Ahmed et al. (2013) \cite{ahmed2013alpha,ahmed2013clusterization}  to describe alpha formation and calculate preformation probabilities for various nuclei \cite{ahmed2017alpha}, the CFM yields the following expression for the alpha preformation probability $P_{\alpha}$ required in the decay width $\Gamma$:
\begin{equation}\label{eq11}
P_{\alpha}=\frac{E_{f\alpha}}{E}.
\end{equation}
In Equation \ref{eq11}, $E_{f\alpha}$ corresponds to the formation energy of the alpha particle, and $E$ is the total energy of the system. Furthermore, the coefficients $E_{f\alpha}$ and $E$ have distinct definitions for even-even, odd-A, and odd-odd nuclei. For even-even nuclei they can be expressed as
\begin{equation}\label{eq12}
\begin{aligned}
E_{f\alpha} = & 3B(A,Z)+B(A-4,Z-2)   \\&
                   -2B(A-1,Z-1)-2B(A-1,Z),
\end{aligned}
\end{equation}
\begin{equation}\label{eq13}
E=B(A,Z)-B(A-4,Z-2).
\end{equation}
where $B(A,Z)$ is the binging energy of the nucleus with the mass number $A$ and proton number $Z$. For the odd-A nuclei they can be written as:
\begin{equation}\label{eq14}
\begin{aligned}
E_{f\alpha}(N_{o},Z_{e}) = & 3B(N_{o}-1,Z_{e})+B(N_{o}-3,Z_{e}-2)  \\&
                   -2B(N_{o}-1,Z_{e}-1)-2B(N_{o}-2,Z_{e}),
\end{aligned}
\end{equation}
\begin{equation}\label{eq15}
E(N_{o},Z_{e})=B(N_{o},Z_{e})-B(N_{o}-3,Z_{e}-2).
\end{equation}
\begin{equation}\label{eq16}
\begin{aligned}
E_{f\alpha}(N_{e},Z_{o}) = & 3B(N_{e},Z_{o}-1)+B(N_{e}-2,Z_{o}-3)  \\&
                   -2B(N_{e},Z_{o}-2)-2B(N_{e}-1,Z_{o}-1),
\end{aligned}
\end{equation}
\begin{equation}\label{eq17}
E(N_{e},Z_{o})=B(N_{e},Z_{o})-B(N_{e}-2,Z_{o}-2).
\end{equation}

For the odd-odd nuclei they can be shown as:
\begin{equation}\label{eq18}
\begin{aligned}
E_{f\alpha}(N_{o},Z_{o}) = & 3B(N_{o}-1,Z_{o}-1)+B(N_{o}-3,Z_{o}-3)  \\&
                   -2B(N_{o}-1,Z_{o}-2)-2B(N_{o}-2,Z_{o}-1),
\end{aligned}
\end{equation}
\begin{equation}\label{eq19}
E(N_{o},Z_{o})=B(N_{o},Z_{o})-B(N_{o}-3,Z_{o}-3).
\end{equation}
\par{}
To overcome the limitations of the Royer formula for suppressed $\alpha$-decay half-lives, Deng et al. extended \cite{deng2020improved} it by including contributions from the centrifugal potential and the blocking effect of unpaired nucleons, leading to a new expression (DZR):
\begin{equation}\label{eq20}
log_{10}T_{1/2} = a + bA^{1/6}\sqrt{Z}+c\frac{Z}{\sqrt{Q_{\alpha}}}+dl(l+1)+h.
\end{equation}
where $A$, $Z$, $Q_{\alpha}$ and $l$ denotes the mass number, proton number, $\alpha$-decay energy of parent nuclei and the angular momentum of the emitted $\alpha$ particle. Based on the results reported by Deng et al., the parameters are given as $a = -26.8125$, $b = -1.1255$, $c = -1.6057$ and $d = 0.0513$. It should be noted that the original work does not provide the corresponding uncertainties for these parameters. Meanwhiles, the values of $h$ for different $\alpha$-decay cases are shown as
\begin{equation}\label{eq21}
h= \left\{\begin{matrix}
 0, &for  &even-even    \\
 0.2821,& for &even-odd  \\
0.3625,&  for&  odd-even \\
  0.7486,&  for&  odd-even \\
\end{matrix}\right.
\end{equation} 
\par{}
An improved model, the Modified Universal Decay Law (MUDL), was developed by Soylu et al \cite{soylu2021extended}. This was achieved by extending the original UDL to include dependencies on angular momentum and isospin. It can be shown as:
\begin{equation}\label{eq22}
\begin{aligned}
log_{10}T_{1/2} = &aZ_{c}Z_{d}\sqrt{\frac{A}{Q_{c}}} + b\sqrt{AZ_{c}Z_{d}*(A_{d}^{1/3}+A_{c}^{1/3})}\\&
                  +c+d\sqrt{AZ_{c}Z_{d}*(A_{d}^{1/3}+A_{c}^{1/3})}\sqrt{l(l+1)}.
\end{aligned}
\end{equation}
where A can be expressed $A=A_{c}A_{d}/(A_{c}+A_{d})$. $A_{c}$ and $A_{d}$ are the cluster and daughter nucleus. By fitting the experimental data of $\alpha$-decay half-lives, the vaule of constants $a =0.4392060\pm0.00398$, $b =-0.3944174\pm0.009317$, $c = -27.0648730\pm0.697731$ and $d =0.0051825\pm0.00042$
 \subsection{Nonlocality Effect}\label{B2}
\par{}
The nonlocality effect of the alpha particle was introduced into $\alpha$-decay theory in the pioneering work of E. L. Medeiros \cite{medeiros2022nonlocality}. Following this approach, the reduced mass of the $\alpha$-daughter nucleus system can be redefined as:
\begin{equation}\label{eq20}
\mu=\frac{m^{*}M}{m^{*}+M},
\end{equation}
where $M$ denotes the nuclear mass of the daughter nucleus.R.A. Zureikat and M.I. Jaghoub derived a coordinate-dependent  mass $m^{*}$ from the gradient of the velocity-dependent potential. This mass $m^{*}$ can be written in the form:
\begin{equation}\label{eq21}
m^{*}=\frac{m}{1-\rho (r)},
\end{equation}
where $m$ is the free mass of $\alpha$ nucles. R.A. Zureikat and M.I. Jaghoub \cite{jaghoub2011novel, zureikat2013surface,alameer2021nucleon} propose that $\rho (r)$  is an isotropic function of the radial variable $r$, which is explicitly given by the reciprocal of the gradient of $[ 1 +\mathrm{exp}(\frac{r - R_{S}}{a_{S}})]^{-1}$:
\begin{equation}\label{eq22}
\rho(r)=\rho_{S}a_{s}\frac{\mathrm{d}}{\mathrm{d}r}\left [ 1 +\mathrm{exp}(\frac{r - R_{S}}{a_{S}}) \right ]^{-1},
\end{equation}
We retain the parameterization established in the earlier study by  E. L. Medeiros: the $R_{S}$ parameter is defined as $R_{S} = R + \Delta R$ ($\Delta R = 3.44$ fm). Here, $a_{s}$ represents the diffuseness of the alpha-core nuclear potential, as defined in the study by E. L. Medeiros. Hence, its value is taken to be identical to that of $a$ in Equation \ref{eq6}.
\begin{figure*}[ht]
    \centering
    \includegraphics[width=\textwidth]{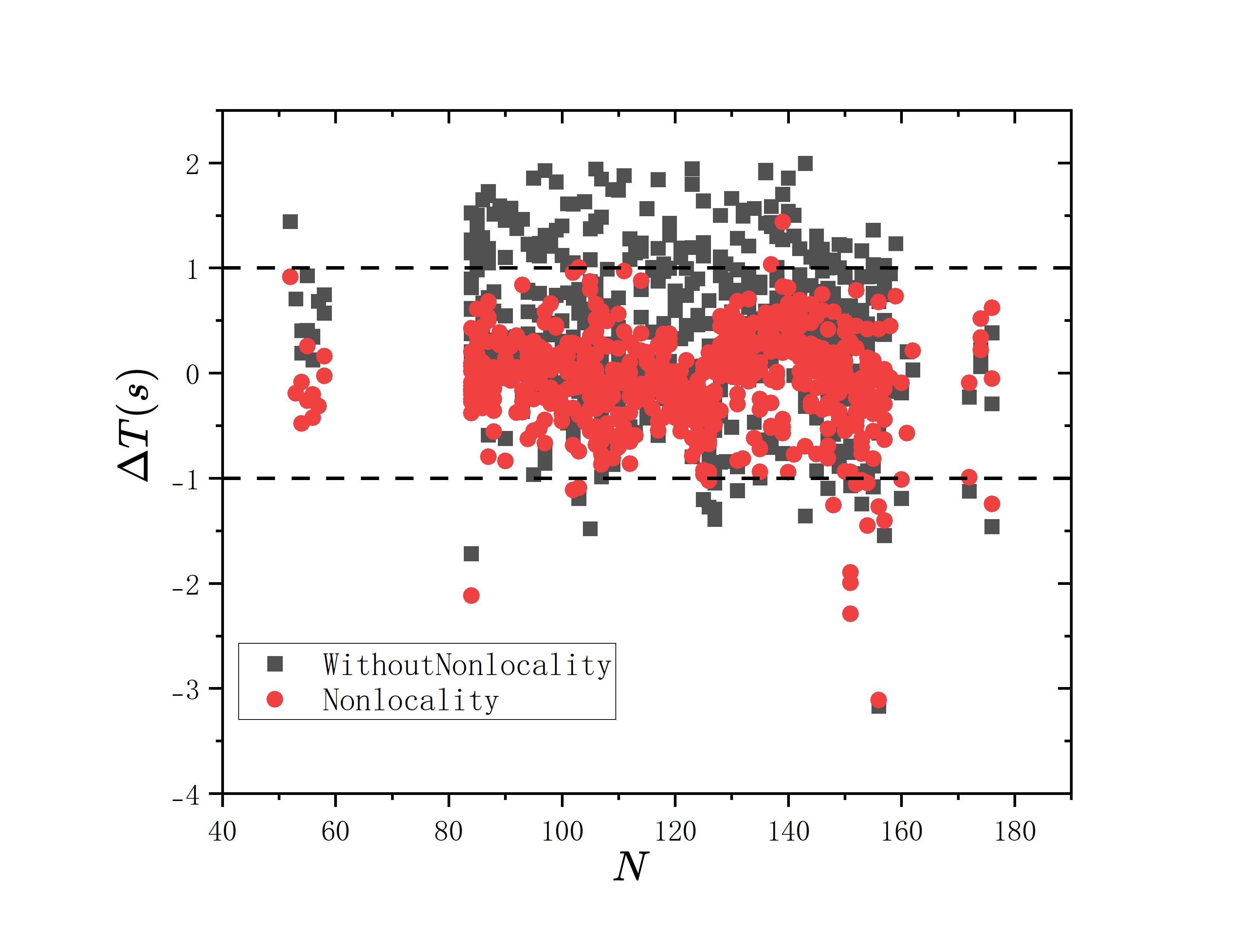}
    \caption{ The abscissa is neutron number $N$ and the ordinate is the value of  $\mathrm{log_{10}}(T_{1/2}^{\mathrm{cal}}/T_{1/2}^{\mathrm{exp}})$. The black squares and red dots represent the theoretical calculation results obtained by using the TPA that takes into account the mass parameters of the nonlocality effect optimized by XGBRegressor model.}
    \label{imag1}
\end{figure*}
\begin{figure*}[ht]
    \centering
    \includegraphics[width=\textwidth]{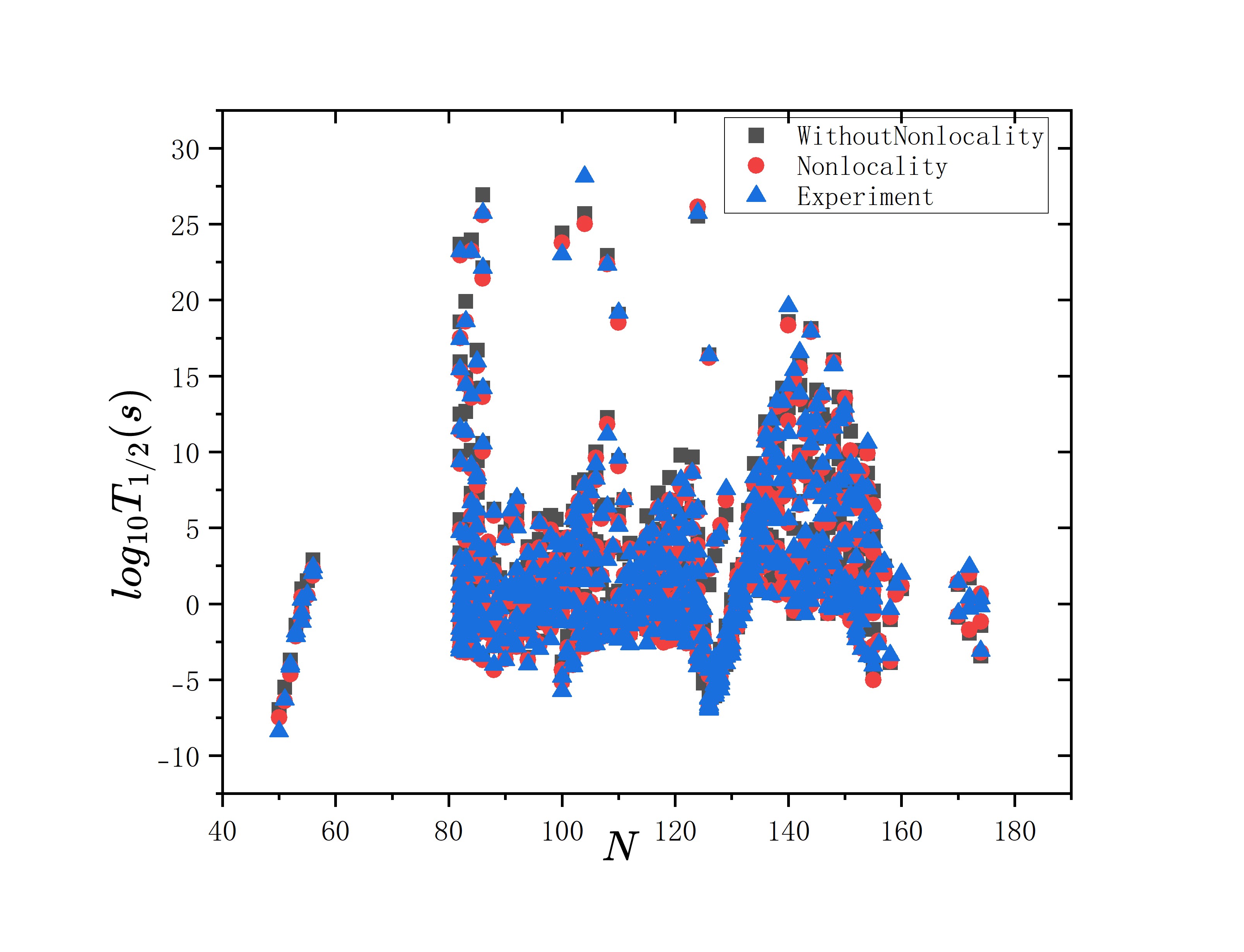}
    \caption{ The abscissa is neutron number $N$ and the ordinate is the value of  $log_{10}T_{1/2}$. The black squares  in left column, red dots in left column, and blue triangle in left column represent the theoretical calculation results obtained by using the TPA that takes into account the mass parameters of the nonlocality effect optimized by the XGBoost regression model(XG), the theoretical calculation results obtained by using the TPA without nonlocality effect, and experiment of $\alpha$-decay half-live, respectively.}
    \label{imag2}
\end{figure*}
\subsection{Machine learning methods}\label{C3}
\par{}
While many linear problems can be solved either analytically or numerically, most nonlinear problems-beyond those involving specific functions like exponentials, logarithms, or sines-pose significant computational challenges. Machine learning (ML) serves as a powerful tool for tackling such nonlinearities. It operates by learning patterns from large datasets to make predictions. Specifically, an ML model typically partitions data into training and test sets. Through an iterative process, it minimizes the discrepancy between its predictions and the true values in the training set. Crucially, to ensure generalization, the model's performance must be evaluated on both the training and test sets. Thus, the primary tasks of ML are prediction and regression. This work focuses on the prediction task.
\par{}
Among classical ML methods, such as random forests \cite{breiman2001random}, support vector machines, decision trees, and eXtreme Gradient Boosting(XG) \cite{chen2016xgboost} , we adopt the gradient boosting framework based on our prior related work. This algorithm is built on a gradient-boosting foundation. It is designed with gradient-enhanced decision trees to improve performance and supports parallel computing, which significantly accelerates model training. This efficiency is a key advantage for processing extensive alpha-decay datasets in subsequent calculations.
\begin{table*}[ht]
    \renewcommand{\arraystretch}{1}
    \setlength{\tabcolsep}{0.4cm}
    \centering
    \caption{The standard deviations between the experimental $\alpha$ decay half-lives and calculated ones with improved TPA by considering nonlocality effect in alpha decay using XG to optimize the mass parameter $\rho_{S}$. }
    \begin{ruledtabular}
    \scalebox{1}{
    \begin{tabular}{ccc}
        Nucleus&XG&Without Nonlocality effect\\
        \colrule
        $\sigma$   &0.46929&0.82058\\
    \end{tabular}
    }
    \label{tab1}
    \end{ruledtabular}
\end{table*}
\begin{table*}[ht]
    \renewcommand{\arraystretch}{1}
    \setlength{\tabcolsep}{0.4cm}
    \centering
    \caption{The mass parameters $\rho_{S}$ adjustments for  three sets nucleus such as even-even nucleus, odd-A nucleus and odd-odd nucleus. The $\rho_{S}$ is adjusted to minimize the differences between the experimental and calculated half-lives for corresponding nuclei. lg$_{1/2}^{\mathrm{XG}}$, lg$_{1/2}^{\mathrm{TPA}}$ and lg$_{1/2}^{\mathrm{exp}}$  are the logarithms of the corresponding calculated half-life, in s. The experimental $\alpha$ decay half-lives are taken form the nuclear properties table NUBASE2020 \cite{kondev2021nubase2020}. The $\alpha$-decay energy are obtained by WS4 \cite{wang2014surface}}
    \begin{ruledtabular}
    \scalebox{1}{
    \begin{tabular}{cccccc}
        Case& Nucleus& $\rho_{S}$ & $\mathrm{lg}_{1/2}^{\mathrm{XG}}$&$\mathrm{lg}_{1/2}^{\mathrm{TPA}}$&$\mathrm{lg}_{1/2}^{\mathrm{exp}}$\\
        \colrule
        Even-Even&$^{110}54$ &2.477515& -1.091 &-0.4979 &-0.84 \\
        &$^{114}56$ & 2.1545 & 1.853 &2.412 &1.694\\
       &$^{184}78$ & 0.9865 & 7.800 & 8.1722 &7.768\\
       &$^{270}110$ &0.5542 & -3.780 & -3.8754 &-3.688\\
       Even-Odd&$^{105}52$ &-0.2682 & -6.3908&-5.4948 &-6.1986 \\
        &$^{167}76$ & 5.400 & 0.4677 &1.7294 &0.2162\\
       &$^{175}80$ & 14.711 & -1.9148 & -8.6536 &-1.9914\\
       &$^{267}110$ &51.190 & -5.0101 & -4.2531 &-5\\
    Odd-Even&$^{237}93$ &1.4807 & 13.5129&14.3819 &13.8303 \\
        &$^{249}97$ & 0.4026 &12.2142&12.7654 &12.29\\
       &$^{245}101$ & 3.2360 & -0.3993& -0.06969 &-0.4202\\
       &$^{261}106$ &2.790324 & -1.008& 0.0294 &-0.7292\\
    Odd-Odd&$^{206}85$ &-1.6879 & 5.2180 &4.8260 &5.3096 \\
        &$^{212}85$ & -0.70 &-0.6734&-1.7978 &-0.5031\\
       &$^{228}91$ & -0.1762 & 6.3474& 5.9234 &6.6316\\
       &$^{244}101$ &-0.5694 & -0.4681& -0.6461 &-0.4437\\
    \end{tabular}
    }
    \label{tab2}
    \end{ruledtabular}
\end{table*}
\begin{table*}[ht]
    \renewcommand{\arraystretch}{1}
    \setlength{\tabcolsep}{0.4cm}
    \centering
    \caption{Summary of Model Evaluation Parameters. }
    \begin{ruledtabular}
    \scalebox{1}{
    \begin{tabular}{cccc}
        Model&the mean squared error(MSE)&root mean squared error (RMSE)&coefficient of determination($R^{2}$)\\
        \colrule
        $XG$   &$2.68\pm 1.43$&$1.60\pm0.33$&$0.56\pm0.18$\\
    \end{tabular}
    }
    \label{tab3}
    \end{ruledtabular}
\end{table*}
\begin{table*}[ht]
  \caption{The $\alpha$-decay half-lives of 142 even-even, odd-A and odd-odd nuclei with Z = 117-120 predicted by the improved TPA, the modified universal decay law (MUDL), and the unitary Royer formula (DZR). The $\alpha$-decay energy $Q_{\alpha}$ are taken from WS4 \cite{wang2014surface}}
  \label{tab4}
  \begin{tabularx}{\textwidth}{XXXXX}
    \toprule
    Z & N & Thiswork & MUDL & DZR \\
    \midrule

117 & 162 & -5.58185 & -4.59531 & -5.68754 \\
117 & 163 & -4.08662 & -4.16646 & -4.82268 \\
117 & 164 & -4.57336 & -3.88172 & -5.0276  \\
117 & 165 & -3.45859 & -3.46336 & -4.17256 \\
117 & 166 & -3.84259 & -2.92039 & -4.13565 \\
117 & 167 & -2.70850 & -2.52048 & -3.29790 \\
117 & 168 & -2.91560 & -2.02485 & -3.30533 \\
117 & 169 & -2.17723 & -1.87056 & -2.69761 \\
117 & 170 & -2.70572 & -1.89758 & -3.19445 \\
117 & 171 & -1.86511 & -1.55488 & -2.41027 \\
117 & 172 & -2.23602 & -1.27078 & -2.61578 \\
117 & 173 & -1.77023 & -1.37070 & -2.24603 \\
117 & 174 & -2.33870 & -1.06034 & -2.42694 \\
117 & 175 & -0.90675 & -0.44243 & -1.38507 \\
117 & 176 & -1.21124 & -0.24603 & -1.67268 \\
117 & 177 & -1.35954 & -0.80210 & -1.72995 \\
117 & 178 & -2.05641 & -1.11374 & -2.49316 \\
117 & 179 & -1.37832 & -0.87500 & -1.80627 \\
117 & 180 & -1.75452 & -0.73438 & -2.14604 \\
117 & 181 & -1.54098 & -0.98360 & -1.91594 \\
117 & 182 & -2.16243 & -1.15892 & -2.55145 \\
117 & 183 & -3.60263 & -3.13976 & -3.71092 \\
117 & 184 & -4.74759 & -3.94720 & -5.16943 \\
117 & 185 & -4.11981 & -3.90111 & -4.43100 \\
117 & 186 & -3.52524 & -2.36953 & -3.70040 \\
117 & 187 & -2.17151 & -1.67228 & -2.35208 \\
117 & 188 & -4.20114 & 0.12296  & -1.14282 \\
117 & 189 & 0.59792  & 1.13081  & 0.75957  \\
117 & 190 & 1.90837  & 3.72663  & 1.86854  \\
117 & 191 & 2.95970  & 4.08374  & 2.78762  \\
117 & 192 & 3.34604  & 5.13328  & 3.29858  \\
117 & 193 & 5.06353  & 6.43925  & 4.98429  \\
117 & 194 & 6.97894  & 8.67180  & 7.33253  \\
117 & 195 & 7.69556  & 9.24273  & 7.83377  \\
117 & 196 & 8.55302  & 10.85812 & 8.58749  \\
117 & 197 & 9.45770  & 10.66841 & 8.87279  \\
117 & 198 & 8.51868  & 10.76879 & 8.49479  \\
118 & 162 & -6.32522 & -5.95911 & -6.53774 \\
118 & 163 & -3.81983 & -2.26401 & -2.98699 \\
118 & 164 & -5.55231 & -5.40372 & -6.02499 \\
118 & 165 & -5.22610 & -5.07915 & -5.36222 \\
118 & 166 & -4.99684 & -4.74820 & -5.41848 \\
118 & 167 & -4.84747 & -4.50845 & -4.83513 \\
118 & 168 & -4.30894 & -4.10720 & -4.82557 \\
118 & 169 & -4.58989 & -4.06827 & -4.43024 \\
118 & 170 & -4.36904 & -4.10420 & -4.82998 \\
118 & 171 & -3.73043 & -2.60192 & -3.78886 \\
118 & 172 & -3.65194 & -3.29983 & -4.08411 \\
118 & 173 & -3.49842 & -2.21649 & -3.43620 \\
118 & 174 & -3.51286 & -3.23301 & -4.02871 \\
118 & 175 & -3.14222 & -2.53213 & -3.01366 \\
118 & 176 & -2.46784 & -2.17287 & -3.04341 \\
118 & 177 & -3.58346 & -3.05575 & -3.51090 \\
118 & 178 & -3.59022 & -3.26284 & -4.07077 \\
118 & 179 & -3.21474 & -1.84144 & -3.10935 \\
118 & 180 & -3.11713 & -2.74581 & -3.59381 \\
118 & 181 & -3.24139 & -1.80724 & -3.08530 \\
118 & 182 & -3.38833 & -2.98588 & -3.82547 \\
118 & 183 & -3.50595 & -2.30263 & -3.69558 \\
118 & 184 & -5.68566 & -5.48150 & -6.16810 \\
118 & 185 & -4.90679 & -3.92196 & -5.08026 \\
118 & 186 & -4.52285 & -4.08833 & -4.87100 \\
118 & 187 & -2.60883 & -1.67047 & -2.98088 \\
118 & 188 & -1.85976 & -0.93666 & -1.92828 \\
    \bottomrule
  \end{tabularx}
\end{table*}
\begin{table*}[ht]
  \caption{continued.}
  \label{tab5}
  \begin{tabularx}{\textwidth}{XXXXX}
    \toprule
    Z & N & Thiswork & MUDL & DZR \\
    \midrule
118 & 189 & -0.09157 & 1.51463 & -0.00793 \\
118 & 190 & 0.43131 & 1.23232 & 0.09463 \\
118 & 191 & 1.46087 & 3.35217 & 1.70378 \\
118 & 192 & 2.26935 & 3.33115 & 2.05177 \\
118 & 193 & 6.22297 & 7.36164 & 6.18233 \\
118 & 194 & 7.15902 & 8.43857 & 6.82402 \\   
118 & 195 &  7.71016 & 10.35523 &  8.11665 \\
118 & 196 &  6.15897 &  7.44282 &  5.88526 \\
118 & 197 &  6.70285 &  9.30573 &  7.12628 \\
118 & 198 &  6.88769 &  8.25643 &  6.63955 \\
119 & 163 & -3.45345 & -3.79352 & -4.63177 \\
119 & 164 & -5.38726 & -4.79113 & -5.90263 \\
119 & 165 & -4.32645 & -4.43010 & -5.10139 \\
119 & 166 & -4.77962 & -4.10993 & -5.27320 \\
119 & 167 & -4.05456 & -4.01688 & -4.72283 \\
119 & 168 & -4.49063 & -3.87603 & -5.06250 \\
119 & 169 & -3.78834 & -3.69268 & -4.42758 \\
119 & 170 & -4.51219 & -3.66644 & -4.87451 \\
119 & 171 & -3.47723 & -3.36150 & -4.12576 \\
119 & 172 & -3.79173 & -2.95865 & -4.22015 \\
119 & 173 & -3.22766 & -3.00058 & -3.79607 \\
119 & 174 & -3.73519 & -3.08744 & -4.34883 \\
119 & 175 & -2.74114 & -2.45780 & -3.29615 \\
119 & 176 & -3.14050 & -2.35364 & -3.67010 \\
119 & 177 & -3.41516 & -3.03479 & -3.84422 \\
119 & 178 & -3.96100 & -3.16431 & -4.43683 \\
119 & 180 & -3.82622 & -3.53399 & -4.05744 \\
119 & 181 & -3.51137 & -3.54906 & -3.60761 \\
119 & 182 & -3.45482 & -2.42952 & -3.76504 \\
119 & 183 & -3.41241 & -4.07929 & -4.61000 \\
119 & 184 & -5.44197 & -4.68460 & -5.88315 \\
119 & 186 & -4.82300 & -4.43633 & -4.92244 \\
119 & 187 & -2.90121 & -2.76609 & -2.89556 \\
119 & 188 & -1.34099 &  0.56416 & -1.11229 \\
119 & 189 &  1.08391 &  1.89475 &  0.59600 \\
119 & 190 & -0.24872 &  0.59332 & -0.22989 \\
119 & 191 &  1.31269 &  2.27205 &  1.06633 \\
119 & 192 &  4.81209 &  6.13253 &  4.20694 \\
119 & 193 &  5.70476 &  7.10175 &  5.81453 \\
119 & 194 &  6.71452 &  8.34875 &  6.51001 \\
119 & 195 &  7.13650 &  8.73606 &  7.09821 \\
119 & 196 &  5.23598 &  7.14021 &  5.00736 \\
119 & 197 &  4.78087 &  5.82804 &  5.11091 \\
119 & 198 &  7.39252 &  9.34805 &  7.31119 \\
120 & 165 & -5.39408 & -4.05357 & -5.28612 \\
120 & 166 & -5.92385 & -5.83332 & -6.44788 \\
120 & 167 & -6.06783 & -5.82264 & -6.07898 \\
120 & 168 & -5.88926 & -5.80577 & -6.42929 \\
120 & 169 & -5.29709 & -4.31256 & -5.40560 \\
120 & 170 & -5.55539 & -5.35242 & -6.01213 \\
120 & 171 & -5.22488 & -4.11456 & -5.22845 \\
120 & 172 & -5.10753 & -4.90528 & -5.60078 \\
120 & 173 & -5.02437 & -3.87091 & -5.00854 \\
120 & 174 & -5.31013 & -5.15133 & -5.83812 \\
120 & 175 & -5.12493 & -4.74029 & -5.09448 \\
120 & 176 & -4.64049 & -4.45857 & -5.19686 \\
120 & 177 & -5.37899 & -5.02277 & -5.36582 \\
120 & 178 & -5.40527 & -5.16590 & -5.86573 \\
120 & 179 & -4.60945 & -3.51651 & -4.70092 \\
120 & 180 & -4.60787 & -4.26456 & -5.02927 \\
120 & 181 & -4.15357 & -2.98434 & -4.21089 \\
    \bottomrule
  \end{tabularx}
\end{table*}
\begin{table*}[ht]
  \caption{continued.}
  \label{tab6}
  \begin{tabularx}{\textwidth}{XXXXX}
    \toprule
    Z & N & Thiswork & MUDL & DZR \\
    \midrule
    120 & 182 & -4.40857 & -4.01154 & -4.79944 \\
    120 & 183 & -4.93157 & -4.04135 & -5.20774 \\
    120 & 184 & -6.42100 & -6.23135 & -6.88319 \\
    120 & 185 & -4.87472 & -3.66076 & -5.00679 \\
    120 & 186 & -5.11406 & -4.53016 & -5.29832 \\
    120 & 187 & -2.92889 & -2.67518 & -3.20377 \\
    120 & 188 & -1.74270 & -1.01168 & -2.01340 \\
    120 & 189 & -0.87006 &  0.91469 & -0.59443 \\
    120 & 190 & -0.93866 & -0.28025 & -1.33599 \\
    120 & 191 & -0.41850 &  1.36045 & -0.18525 \\
    120 & 192 &  0.41484 &  0.98663 & -0.15769 \\
    120 & 193 &  4.20727 &  5.24993 &  4.18946 \\
    120 & 194 &  5.06640 &  6.11169 &  4.62958 \\
    120 & 195 &  2.63445 &  5.09793 &  3.16702 \\
    120 & 196 &  2.52368 &  3.49840 &  2.17818 \\
    120 & 197 &  2.96779 &  5.35955 &  3.40354 \\
    120 & 198 &  3.22243 &  4.31765 &  2.93765 \\
    \bottomrule
  \end{tabularx}
\end{table*}

\section{\label{sec:level3}RESULTS AND DISCUSSION}
\par{}
In this study, an improved Two-Potential Approach (TPA) is developed by incorporating nonlocality effect into the $\alpha$-particle-nucleus interaction kernel. Meanwhile, the XG method is adopted to predict the relevant parameters of the model. The proposed framework is applied to calculate the $\alpha$-decay half-lives of 599 nuclei with 52  $\le $ $\mathbf{Z}$ $\le$ 118 and $\mathbf{N}$ $\ge$ 52.
\par{}
In the XG model, the angular momentum of the parent nucleus, the $\alpha$-preformation probability, the asymmetry term of the parent nucleus, and the square root of the decay energy are taken as input features, while the mass parameter is treated as the output variable. During the training phase, the initial model parameters were manually assigned to ensure that the calculated $\alpha$-decay half-lives deviate from the experimental values by less than 0.001. Subsequently, the built-in automatic optimization routines in Python 3.7 were employed to further optimize the model parameters. Finally, the performance of the model was evaluated using the mean squared error (MSE), the root mean squared error (RMSE), and the coefficient of determination $R^{2}$. As shown in Table \ref{tab3}, the XG model demonstrates a reliable capability in predicting $\alpha$-decay half-lives.
\par{}
In this study, the standard deviation is adopted to quantify the global discrepancy between theoretical and experimental $\alpha$-decay half-lives. In addition, the difference between the natural logarithms of the theoretical and experimental half-lives is used to evaluate the deviation for each nucleus. The corresponding definitions are given as follows:
\begin{equation}\label{eq17}
\sigma =\sqrt{\frac{1}{n}\sum_{i=1}^{n}(\Delta_{i})^{(2)}};\ \Delta_{i}= \mathrm{log_{10}}(T_{i}^{\mathrm{cal}})-\mathrm{log_{10}}(T_{i}^{\mathrm{exp}}),
\end{equation}
$T_{i}^{\mathrm{cal}}$ and $T_{i}^{\mathrm{exp}}$ denote the calculated and experimental $\alpha$-decay half-lives of the i-th nucleus.
\par{}
Figures \ref{imag1} and \ref{imag2} show that, the TPA that accounts for the nonlocality effects of the $\alpha$ particle outperforms the version that omits these effects in fitting the experimental $\alpha$-decay half-life data. Furthermore, Figure \ref{imag2} reveals that when, the $\alpha$-decay half-life reaches a minimum, indicating a pronounced shell effect at this neutron number.

\par{}
From Table \ref{tab1}, it can be seen that the TPA incorporating the nonlocality effects of the $\alpha$ particle exhibits a 74.85$\%$ reduction in RMS compared to the TPA that does not account for these effects. Furthermore, as can be seen from Figures \ref{imag2} and Tables \ref{tab2}, the improved TPA has a half-life for $\alpha$-decay of atomic nuclei with N greater than 80 that is closer to the experimental value of the half-life of atomic nucleus $\alpha$-decay than that of the original TPA.

\par{}
As shown in Table \ref{tab3}, the mean squared error (MSE), root mean squared error (RMSE), and coefficient of determination ($R^{2}$)  are $2.68\pm 1.43$, $1.60\pm0.33$, and $0.56\pm0.18$, respectively. The uncertainties were estimated from the XGB regression model using 50 repetitions of 5-fold cross-validation, providing an assessment of the model's robustness and generalization ability.

\par
\begin{figure*}[b]
    \centering
    \includegraphics[width=\textwidth]{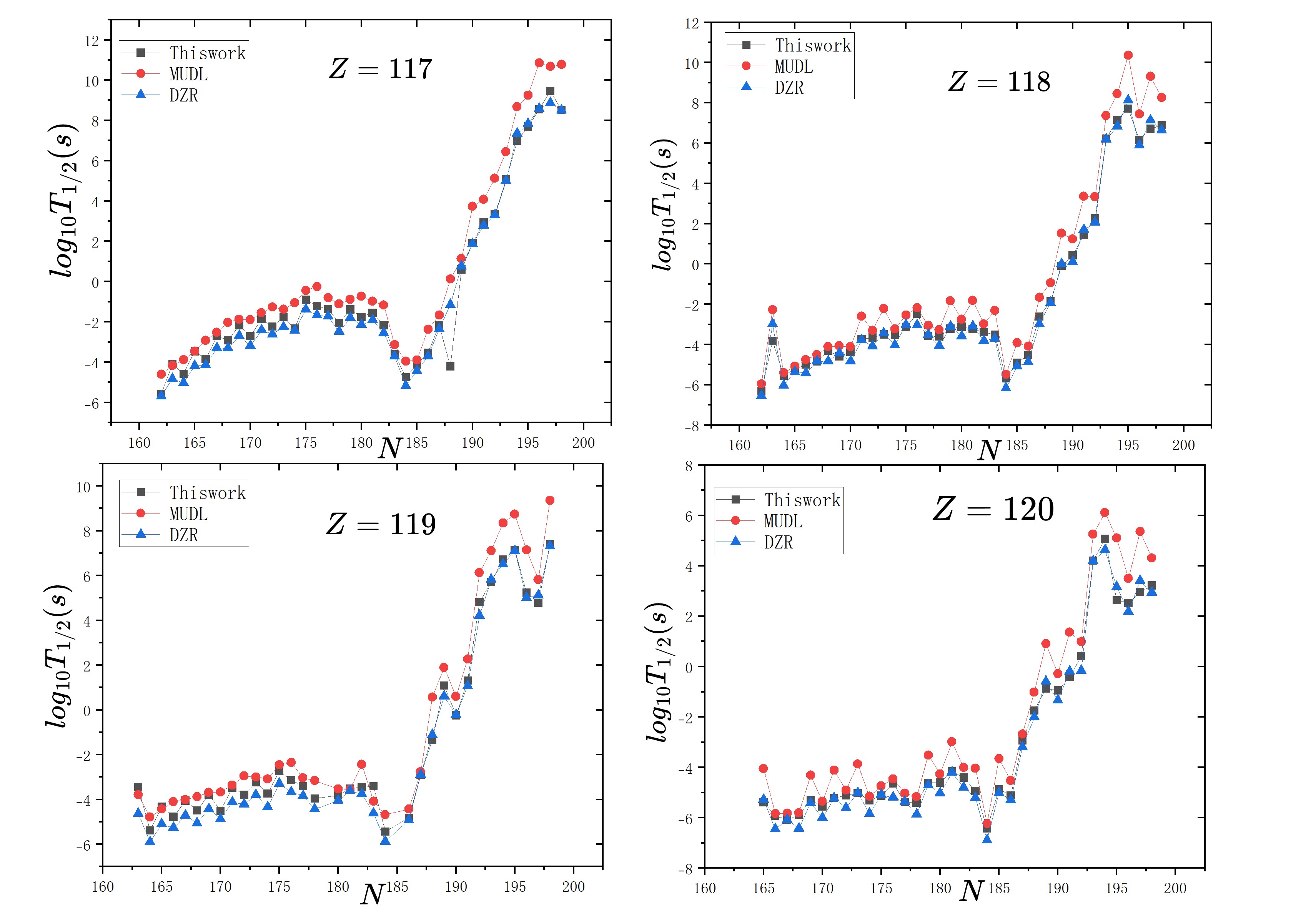}
    \caption{ The calculation values of $\alpha$ decay half-lives for 142 even-even nuclei, odd-A nuclei and odd-odd nuclei of $Z = 117 - 120$  isotopes are presented. The abscissa is neutron number $N$  and the ordinate is logarithm $\mathrm{log_{10}}T_{1/2}$ of calculated half-life in s. The black squares, red dots and blue triangles in the top row denote the calculation results obtained by using the TPA that takes into account the mass parameters of the nonlocality effect optimized by XGBRegressor model, using the MUDL model, and using the DZR model, respectively.The $\alpha$-decay energy $Q_{\alpha}$ are taken from WS4 \cite{wang2014surface}.}
    \label{imag7}
\end{figure*}

{}
In this study, the quality parameters predicted by the XG model were employed to improve the TPA by incorporating the non-local effects of the $\alpha$-particle, with the aim of predicting the $\alpha$-decay half-lives of superheavy nuclei with 117  $\le $ $\mathbf{Z}$ $\le$ 120. The prediction results were then compared with those obtained from the DZR and MUDL models. The results of all three models are presented in Tables \ref{tab4}-\ref{tab6}, where the first and second columns list the proton and neutron numbers of the nuclei, and the third, fourth, and fifth columns correspond to the predictions from this work, MUDL, and DZR, respectively. As shown in Figure \ref{imag7}, our predictions are generally consistent with both MUDL and DZR, and in particular, exhibit a closer agreement with the DZR results. This indicates that our approach is in good agreement with the existing models. Furthermore, Figure \ref{imag7} also reveals that  $\mathbf{N}$  = 184 are the probable neutron magic number, which is consistent with previous studies.

\par{}
The $R^{2}$ value of 0.56 indicates that the current XGB regression model captures only a moderate level of correlation with the true values, suggesting limited predictive accuracy. This limitation may arise from the insufficient inclusion of relevant physical features. In future work, additional physics-informed quantities, such as deformation, could be incorporated to improve the prediction of the mass parameters. Furthermore, as shown in Figure \ref{imag7}, the prediction for the superheavy nucleus with  $Z$ = 117 and $N$ = 188 deviates significantly from the results of the DZR and MUDL models. This discrepancy may be attributed to the reliance of the XG model on the $\alpha$-preformation probability, which is not predicted with sufficient accuracy. In addition, Figure \ref{imag2} indicates that the improved TPA performs worse than the original TPA for certain nuclei, suggesting that the limited predictive capability of the XG model also contributes to these deviations. In future work, we plan to enhance the predictive performance of the XG model or adopt a more advanced machine learning method, as well as employ more accurate calculations of the $\alpha$-preformation probability to improve the reliability of $\alpha$-decay half-life predictions.

\section{\label{sec:level4}SUMMARY AND CONCLUSION}
In this work, we extend the treatment of $\alpha$-decay nonlocal effects to odd-A and odd-odd nuclei within the two-potential approach (TPA) framework. The coordinate-dependent parameters introduced by this extension are predicted using the XGBoost (XG) method. This refined TPA model is applied to calculate the $\alpha$-decay  half-lives of 599 nuclei. Compared to the original TPA, the improved version reduces the root-mean-square (RMS) deviation from experimental data by 74.8$\%$. Finally, we employ the model to predict $\alpha$-decay y half-lives for superheavy nuclei with Z = 117-120. The predictions show excellent agreement with those of the DZR and MUDL models, being nearly identical to the DZR results. These accurate predictions are expected to provide valuable guidance for future searches for superheavy nuclei.

\begin{acknowledgments}
This work was supported by the Zhejiang normal university Doctorial research fund Contract No. ZC302924005.

\end{acknowledgments}

\bibliography{study}

@article{denisov2024empirical,
  title={Empirical relations for $\alpha$-decay half-lives: The effect of deformation of daughter nuclei},
  author={Denisov, V Yu},
  journal={Physical Review C},
  volume={110},
  number={1},
  pages={014604},
  year={2024},
  publisher={APS}
}

@article{gurney1928wave,
  title={Wave mechanics and radioactive disintegration},
  author={Gurney, Ronald W and Condon, Edw U},
  journal={Nature},
  volume={122},
  number={3073},
  pages={439--439},
  year={1928},
  publisher={Nature Publishing Group UK London}
}

@article{gamow1928quantentheorie,
  title={Zur quantentheorie des atomkernes},
  author={Gamow, George},
  journal={Zeitschrift f{\"u}r Physik},
  volume={51},
  number={3},
  pages={204--212},
  year={1928},
  publisher={Springer}
}

@article{matsuse1975study,
  title={Study of the Alpha-Clustering Structure of 20Ne Based on the Resonating Group Method for 16O+ $\alpha$: Analysis of Alpha-Decay Widths and the Exchange Kernel},
  author={Matsuse, Takehiro and Kamimura, Masayasu and Fukushima, Yoshihiro},
  journal={Progress of Theoretical Physics},
  volume={53},
  number={3},
  pages={706--724},
  year={1975},
  publisher={Oxford University Press}
}

@article{wauters1994fine,
  title={Fine structure in the alpha decay of even-even nuclei as an experimental proof for the stability of the Z= 82 magic shell at the very neutron-deficient side},
  author={Wauters, Jan and Bijnens, N and Dendooven, P and Huyse, Marc and Hwang, Han Yull and Reusen, G and von Schwarzenberg, J and Van Duppen, Piet and Kirchner, R and Roeckl, E},
  journal={Physical review letters},
  volume={72},
  number={9},
  pages={1329},
  year={1994},
  publisher={APS}
}

@article{kucuk2020role,
  title={Role of the dynamical polarization potential in explaining the $\alpha$+ 12C system at low energies},
  author={Kucuk, Y and Soylu, Asim and Chamon, LC},
  journal={Nuclear Physics A},
  volume={994},
  pages={121665},
  year={2020},
  publisher={Elsevier}
}

@article{wang2017competition,
  title={Competition between $\alpha$ decay and proton radioactivity of neutron-deficient nuclei},
  author={Wang, YZ and Cui, JP and Zhang, YL and Zhang, S and Gu, JZ},
  journal={Physical Review C},
  volume={95},
  number={1},
  pages={014302},
  year={2017},
  publisher={APS}
}

@article{xiao2020alpha,
  title={$\alpha$-Decay with extremely long half-lives},
  author={Xiao, Yang and Zhang, Shan and Cui, Jianpo and Wang, Yanzhao},
  journal={Indian Journal of Physics},
  volume={94},
  number={4},
  pages={527--533},
  year={2020},
  publisher={Springer}
}

@article{ma2020short,
  title={Short-Lived $\alpha$-Emitting Isotope Np 222 and the Stability of the N= 126 Magic Shell},
  author={Ma, L and Zhang, ZY and Gan, ZG and Zhou, XH and Yang, HB and Huang, MH and Yang, CL and Zhang, MM and Tian, YL and Wang, YS and others},
  journal={Physical Review Letters},
  volume={125},
  number={3},
  pages={032502},
  year={2020},
  publisher={APS}
}

@article{oganessian2011eleven,
  title={Eleven new heaviest isotopes of elements Z= 105 to Z= 117 identified among the products of Bk 249+ Ca 48 reactions},
  author={Oganessian, Yu Ts and Abdullin, F Sh and Bailey, PD and Benker, DE and Bennett, ME and Dmitriev, SN and Ezold, Julie G and Hamilton, JH and Henderson, RA and Itkis, MG and others},
  journal={Physical Review C—Nuclear Physics},
  volume={83},
  number={5},
  pages={054315},
  year={2011},
  publisher={APS}
}

@article{oganessian2004experiments,
  title={Experiments on the synthesis of element 115 in the reaction Am 243 (Ca 48, xn) 115 291- x},
  author={Oganessian, Yu Ts and Utyonkoy, VK and Lobanov, Yu V and Abdullin, F Sh and Polyakov, AN and Shirokovsky, IV and Tsyganov, Yu S and Gulbekian, GG and Bogomolov, SL and Mezentsev, AN and others},
  journal={Physical Review C—Nuclear Physics},
  volume={69},
  number={2},
  pages={021601},
  year={2004},
  publisher={APS}
}

@article{zhang2021new,
  title={New $\alpha$-Emitting Isotope U 214 and Abnormal Enhancement of $\alpha$-Particle Clustering in Lightest Uranium Isotopes},
  author={Zhang, ZY and Yang, HB and Huang, MH and Gan, ZG and Yuan, CX and Qi, C and Andreyev, AN and Liu, ML and Ma, L and Zhang, MM and others},
  journal={Physical Review Letters},
  volume={126},
  number={15},
  pages={152502},
  year={2021},
  publisher={APS}
}

@article{oganessian2000observation,
  title={Observation of the decay of 292 116},
  author={Oganessian, Yu Ts and Utyonkov, VK and Lobanov, Yu V and Abdullin, F Sh and Polyakov, AN and Shirokovsky, IV and Tsyganov, Yu S and Gulbekian, GG and Bogomolov, SL and Gikal, BN and others},
  journal={Physical Review C},
  volume={63},
  number={1},
  pages={011301},
  year={2000},
  publisher={APS}
}

@article{buck1992alpha,
  title={$\alpha$ decay calculations with a realistic potential},
  author={Buck, B and Merchant, AC and Perez, SM},
  journal={Physical Review C},
  volume={45},
  number={5},
  pages={2247},
  year={1992},
  publisher={APS}
}

@article{duarte1998cold,
  title={Cold fission description with constant and varying mass asymmetries},
  author={Duarte, SB and Rodriguez, O and Tavares, OA P and Gon{\c{c}}alves, M and Garc{\'\i}a, F and Guzm{\'a}n, F},
  journal={Physical Review C},
  volume={57},
  number={5},
  pages={2516},
  year={1998},
  publisher={APS}
}

@article{goncalves1993effective,
  title={Effective liquid drop description for the exotic decay of nuclei},
  author={Goncalves, M and Duarte, SB},
  journal={Physical Review C},
  volume={48},
  number={5},
  pages={2409},
  year={1993},
  publisher={APS}
}

@article{basu2003role,
  title={Role of effective interaction in nuclear disintegration processes},
  author={Basu, DN},
  journal={Physics Letters B},
  volume={566},
  number={1-2},
  pages={90--97},
  year={2003},
  publisher={Elsevier}
}

@article{chowdhury2006alpha,
  title={$\alpha$ decay half-lives of new superheavy elements},
  author={Chowdhury, P Roy and Samanta, C and Basu, DN},
  journal={Physical Review C—Nuclear Physics},
  volume={73},
  number={1},
  pages={014612},
  year={2006},
  publisher={APS}
}

@article{gurvitz1987decay,
  title={Decay width and the shift of a quasistationary state},
  author={Gurvitz, SA and Kalbermann, G},
  journal={Physical review letters},
  volume={59},
  number={3},
  pages={262},
  year={1987},
  publisher={APS}
}

@article{buck1993half,
  title={Half-lives of favored alpha decays from nuclear ground states},
  author={Buck, B and Merchant, AC and Perez, SM},
  journal={Atomic Data and Nuclear Data Tables},
  volume={54},
  number={1},
  pages={53--73},
  year={1993},
  publisher={Elsevier}
}

@article{samanta2007predictions,
  title={Predictions of alpha decay half lives of heavy and superheavy elements},
  author={Samanta, C and Chowdhury, P Roy and Basu, DN},
  journal={Nuclear Physics A},
  volume={789},
  number={1-4},
  pages={142--154},
  year={2007},
  publisher={Elsevier}
}

@article{ahmed2017alpha,
  title={Alpha-cluster preformation factor within cluster-formation model for odd-A and odd--odd heavy nuclei},
  author={Ahmed, Saad M Saleh},
  journal={Nuclear Physics A},
  volume={962},
  pages={103--121},
  year={2017},
  publisher={Elsevier}
}

@article{zhang2006alpha,
  title={$\alpha$ decay half-lives of new superheavy nuclei within a generalized liquid drop model},
  author={Zhang, Hongfei and Zuo, Wei and Li, Junqing and Royer, Guy},
  journal={Physical Review C—Nuclear Physics},
  volume={74},
  number={1},
  pages={017304},
  year={2006},
  publisher={APS}
}

@article{viola1966nuclear,
  title={Nuclear systematics of the heavy elements—II Lifetimes for alpha, beta and spontaneous fission decay},
  author={Viola Jr, VE and Seaborg, GT},
  journal={Journal of Inorganic and Nuclear Chemistry},
  volume={28},
  number={3},
  pages={741--761},
  year={1966},
  publisher={Elsevier}
}

@article{qi2009universal,
  title={Universal decay law in charged-particle emission and exotic cluster radioactivity},
  author={Qi, Chong and Xu, FR and Liotta, Roberto J and Wyss, Ramon},
  journal={Physical review letters},
  volume={103},
  number={7},
  pages={072501},
  year={2009},
  publisher={APS}
}

@article{royer2000alpha,
  title={Alpha emission andspontaneous fission through quasi-molecularshapes},
  author={Royer, Guy},
  journal={Journal of Physics G: Nuclear and Particle Physics},
  volume={26},
  number={8},
  pages={1149},
  year={2000},
  publisher={IOP Publishing}
}

@article{oganessian2011synthesis,
  title={Synthesis of the heaviest elements in 48Ca-induced reactions},
  author={Oganessian, Yu Ts},
  journal={Radiochimica Acta},
  volume={99},
  number={7-8},
  pages={429--439},
  year={2011},
  publisher={Oldenbourg Wissenschaftsverlag GmbH}
}

@article{jalili2024decay,
  title={-decay half-life predictions with support vector machine},
  author={Jalili, Amir and Pan, Feng and Draayer, Jerry P and Chen, Ai-Xi and Ren, Zhongzhou},
  journal={Scientific Reports},
  volume={14},
  number={1},
  pages={30776},
  year={2024},
  publisher={Nature Publishing Group UK London}
}

@article{ahmed2013alpha,
  title={Alpha-cluster preformation factors in alpha decay for even--even heavy nuclei using the cluster-formation model},
  author={Ahmed, Saad M Saleh and Yahaya, Redzuwan and Radiman, Shahidan and Yasir, Muhamad Samudi},
  journal={Journal of Physics G: Nuclear and Particle Physics},
  volume={40},
  number={6},
  pages={065105},
  year={2013},
  publisher={IOP Publishing}
}

@article{ahmed2013clusterization,
  title={Clusterization probability in alpha-decay 212Po nucleus within cluster-formation model; a new approach},
  author={Ahmed, Saad M Saleh and Yahaya, Redzuwan and Radiman, Shahidan},
  journal={Romanian Reports in Physics},
  volume={65},
  number={4},
  pages={1281--1300},
  year={2013}
}

@article{sun2016systematic,
  title={Systematic study of $\alpha$ decay half-lives for even-even nuclei within a two-potential approach},
  author={Sun, Xiao-Dong and Guo, Ping and Li, Xiao-Hua},
  journal={Physical Review C},
  volume={93},
  number={3},
  pages={034316},
  year={2016},
  publisher={APS}
}

@article{sun2017systematic,
  title={Systematic study of $\alpha$ decay for odd-A nuclei within a two-potential approach},
  author={Sun, Xiao-Dong and Duan, Chao and Deng, Jun-Gang and Guo, Ping and Li, Xiao-Hua},
  journal={Physical Review C},
  volume={95},
  number={1},
  pages={014319},
  year={2017},
  publisher={APS}
}

@article{jaghoub2011novel,
  title={Novel source of nonlocality in the optical model},
  author={Jaghoub, MI and Hassan, MF and Rawitscher, GH},
  journal={Physical Review C—Nuclear Physics},
  volume={84},
  number={3},
  pages={034618},
  year={2011},
  publisher={APS}
}

@article{zureikat2013surface,
  title={Surface and volume term nonlocalities in the proton--nucleus elastic scattering process},
  author={Zureikat, RA and Jaghoub, MI},
  journal={Nuclear Physics A},
  volume={916},
  pages={183--209},
  year={2013},
  publisher={Elsevier}
}

@article{alameer2021nucleon,
  title={Nucleon-nucleus velocity-dependent optical model: Revisited},
  author={Alameer, Sajedah and Jaghoub, MI and Ghabar, I},
  journal={Journal of Physics G: Nuclear and Particle Physics},
  volume={49},
  number={1},
  pages={015106},
  year={2021},
  publisher={IOP Publishing}
}

@article{breiman2001random,
  title={Random forests},
  author={Breiman, Leo},
  journal={Machine learning},
  volume={45},
  pages={5--32},
  year={2001},
  publisher={Springer}
}

@inproceedings{chen2016xgboost,
  title={Xgboost: A scalable tree boosting system},
  author={Chen, Tianqi and Guestrin, Carlos},
  booktitle={Proceedings of the 22nd acm sigkdd international conference on knowledge discovery and data mining},
  pages={785--794},
  year={2016}
}

@article{medeiros2022nonlocality,
  title={Nonlocality effect in $\alpha$ decay of heavy and superheavy nuclei},
  author={Medeiros, Emil L and Teruya, N and Duarte, S{\'e}rgio B and Tavares, OAP},
  journal={Physical Review C},
  volume={106},
  number={2},
  pages={024608},
  year={2022},
  publisher={APS}
}

@article{soylu2021extended,
  title={Extended universal decay law formula for the $\alpha$ and cluster decays},
  author={Soylu, As{\i}m and Qi, Chong},
  journal={Nuclear Physics A},
  volume={1013},
  pages={122221},
  year={2021},
  publisher={Elsevier}
}

@article{akrawy2018new,
  title={New empirical formula for $\alpha$-decay calculations},
  author={Akrawy, Dashty T and Ahmed, Ali H},
  journal={International Journal of Modern Physics E},
  volume={27},
  number={08},
  pages={1850068},
  year={2018},
  publisher={World Scientific}
}

@article{yang2026alpha,
  title={$\alpha$-decay half-lives of superheavy nuclei with support-vector regression},
  author={Yang, Haitao and Li, Xiaopan and Song, Xiefei and Ma, Dianxu and Yu, Gongming and Bao, Xiaojun},
  journal={Physical Review C},
  volume={113},
  number={1},
  pages={014307},
  year={2026},
  publisher={APS}
}

@article{hu2025nonlocality,
  title={Nonlocality Effect in the Tunneling of Alpha Radioactivity with the Aid of Machine Learning},
  author={Hu, Jinyu and Wu, Chen},
  journal={arXiv preprint arXiv:2510.18199},
  year={2025}
}

@article{deng2020improved,
  title={Improved empirical formula for $\alpha$-decay half-lives},
  author={Deng, Jun-Gang and Zhang, Hong-Fei and Royer, Guy},
  journal={Physical Review C},
  volume={101},
  number={3},
  pages={034307},
  year={2020},
  publisher={APS}
}

@article{kondev2021nubase2020,
  title={The NUBASE2020 evaluation of nuclear physics properties},
  author={Kondev, FG and Wang, Meng and Huang, WJ and Naimi, S and Audi, G},
  journal={Chinese Physics C},
  volume={45},
  number={3},
  pages={030001},
  year={2021},
  publisher={IOP Publishing}
}

@article{wang2014surface,
  title={Surface diffuseness correction in global mass formula},
  author={Wang, Ning and Liu, Min and Wu, Xizhen and Meng, Jie},
  journal={Physics Letters B},
  volume={734},
  pages={215--219},
  year={2014},
  publisher={Elsevier}
}
\end{document}